\documentclass[useAMS,usenatbib]{mn2e}
\usepackage{graphicx}
\usepackage{amssymb}
\usepackage{amsmath}
\usepackage{algpseudocode}
\usepackage{multicol}
\usepackage{longtable}
\usepackage{algorithm}
\usepackage{cprotect}
\title[]{CH$^+$ depletion by atomic hydrogen: accuracy of new rates in photo-dominated and self-shielded environments}

\author[S. Bovino et al.]{ S. Bovino$^1$, T. Grassi$^2$, M. Tacconi$^3$, and F. A. Gianturco$^{2,4,5}$\thanks{Corresponding author: francesco.gianturco@uniroma1.it}\\
$^1$Institut f\"ur Astrophysik Georg-August-Universit\"at, Friedrich-Hund Platz 1, 37077 G\"ottingen, Germany\\
$^2$Department of Chemistry, Sapienza University of Rome, P.le A. Moro, 00185, Rome, Italy\\
$^3$CINECA, via Magnanelli 6/3, 40033 Casalecchio di Reno, Bologna, Italy\\
$^4$Institute of Ion Physics, University of Innsbruck, Technikerstrasse 25, 6020, Innsbruck, Austria\\
$^5$Scuola Normale Superiore, Piazza de' Cavalieri, 56125, Pisa, Italy}

\begin{document}
\newcommand{\ith}{$i$th }
\newcommand{\jth}{$j$th }
\newcommand{\nth}{$n$th }

\newcommand{\dd}{\mathrm d}
\newcommand{\mA}{\mathrm A}
\newcommand{\mB}{\mathrm B}
\newcommand{\mC}{\mathrm C}
\newcommand{\mD}{\mathrm D}
\newcommand{\mE}{\mathrm E}
\newcommand{\mH}{\mathrm H}
\newcommand{\mSi}{\mathrm Si}
\newcommand{\mO}{\mathrm O}
\newcommand{\cmc}{\mathrm{cm}^{-3}}
\newcommand{\real}{\mathbb R}
\newcommand{\superscript}[1]{\ensuremath{^{\scriptscriptstyle\textrm{#1}\,}}}
\newcommand{\trader}{\superscript{\textregistered}}

\newcommand\mnras{MNRAS}
\newcommand\apj{ApJ}
\newcommand\aap{A\&A}
\newcommand\apss{Ap\&SS}

\date{Accepted *****. Received *****; in original form ******}

\pagerange{\pageref{firstpage}--\pageref{lastpage}} \pubyear{2013}

\maketitle

\label{firstpage}
 
  \begin{abstract}
   A detailed quantum analysis of a ionic reaction with a  crucial role in the ISM is carried out to generate ab initio reactive cross sections with a quantum method. From them we obtain the corresponding CH$^+$ depletion rates over a broad range of temperatures. The new rates are further linked to a complex chemical network that shows the evolution in time of the CH$^+$ abundance in photodissociation region (PDR) and molecular cloud (MC) environments.
  The evolutionary abundances of CH$^+$ are given by numerical solutions of a large set of coupled, first-order kinetics equations by employing the new chemical package \textsc{krome}.
The differences found between all existing calculations from low-T experiments are explained via a simple numerical model that links the low-T cross section reductions to collinear approaches where nonadiabatic crossings dominate. The analysis of evolutionary abundance of CH$^+$ reveals that the important region for the depletion reaction of this study is that above 100 K, hence showing that, at least for this reaction, the differences with the existing low-temperature experiments are of essentially no  importance within  the astrochemical environments. A detailed analysis of the chemical network involving CH$^+$ also shows that a slight decrease in the initial oxygen abundance might lead to higher CH$^+$ abundance since the main chemical carbon ion depletion channel is reduced in efficiency. This simplified observation might provide an alternative starting point to understand the problem of astrochemical models in matching the observed CH$^+$ abundances.
\end{abstract}


%---------------------------------------------------------------------------
%---------------------------------------------------------------------------
%---------------------------------------------------------------------------
\section{Introduction}
The methylidine cation CH$^+$ was observed a while ago for the first time \citep{Douglas1941} in the diffuse interstellar medium (ISM)  and was followed by further detections in a variety of interstellar and circumstellar environments. In its earlier detections, its A$^1\Pi\leftarrow X^1\Sigma^+$ electronic band system was observed \citep{Crane1995,Weselak2008}, while the more recent data from the \textit{Infrared Space Observatory}\footnote{http://iso.esac.esa.int} \citep{Kessler1996} and from the \textit{Herschel Space Telescope}\footnote{http://herschel.esac.esa.int/} \citep{Pillbratt2010} have given access to the far infrared (FIR) spectrum of this molecule in several remote star forming regions \citep{Falgarone2010}. The fairly large number of subsequent detections shows that the presence of CH$^+$ can be considered as confirmed throughout the interstellar matter \citep{Godard2013}.

Despite the large effort made by observers to detect with great accuracy the CH$^+$ bands, this molecule still remains a puzzle in the modern molecular astrophysics. The observed averaged abundances, in fact, are still orders of magnitudes larger compared to the one provided by astrochemical models, even if some attempts to solve this puzzle came out during the last years (see discussion below). 

The methylidine has also been known to play a significant role in various steps of the complex chemical network of those molecular processes and reactions which are taken to occur in interstellar and circumstellar regions. For example, its hydrogenation reaction is involved in two important species in that network: the methyledene ion CH$_2^+$ and the methyl cation CH$_3^+$. Although the latter molecules are found to be rapidly destroyed by efficient dissociative recombination with electrons via
\begin{equation}
	\rm CH_3^+ + e^- \rightarrow CH + H_2
\end{equation}
\begin{equation}
	\rm CH_2^+ + e^- \rightarrow C + H_2\,,
\end{equation}
they can also react with oxygen and nitrogen atoms forming in addition CO$^+$, CN$^+$, HCO$^+$, and HCN$^+$ plus other cation precursors of CO, HCN, etc. \citep{Godard2013}, thus expected to induce a departure of carbon from its ionisation equilibrium \citep{Godard2009}. CH$^+$ therefore initiates an extensive chemical chain of processes which evolve into the formation of more complex species.

Since CH$^+$ is chemically a very reactive ion, its reaction path to destruction by hydrogen abstraction has been considered, together with dissociative recombination and the above hydrogenation processes, one of the important ways to its destruction \citep{McEwan1999,Larsson2008,Mitchell1990}: it is thus both important and interesting to assess as accurately as possible the actual efficiency of that destruction path

\begin{equation}\label{eq:reaction}
	\rm CH^+ + H \rightarrow C^+ + H_2\,,
\end{equation}
so that one may link the above reaction with the additional variety of chemical processes which are important to model the chemistry of CH$^+$ in various ISM environments \citep{Godard2013}. 

In the last few years there have been a series of papers which have dealt first with its photon-induced formation:
\begin{equation}
	\rm{ C^+(^2P_{3/2,1/2}) + H(^2S_{1/2}) \rightarrow CH^+(^1\Sigma^+) + }h\nu
\end{equation}
thereby producing the relevant radiative association rates \citep{Barinovs2006}, and also with the experimental, low-temperature study of its destruction reaction by Eqn.(\ref{eq:reaction}) with slow H atoms \citep{Plasil2011}. 

More recently, the formation reaction of CH$^+$ has been analysed with new computational data which start from a vibrationally excited ($\nu$ = 1) H$_2$ partner of C$^+$ and investigate the high-T regimes using a quantum wave-packet method \citep{Zanchet2013}. 

The latter formation reaction is thought to be the main path of CH$^+$ formation even if a series of other physical processes have to be invoked to excite the H$_2$ molecules into higher vibrational levels in order to overcome the endothermicity of the reaction (4177 K). For instance, neutral shocks \citep[e.g.]{Elitzur1978}, and magnetohydrodynamic (MHD) shocks \citep[e.g]{Draine1986} have been suggested earlier but seem to be both ruled out by observations \citep{Gredel1993,Crawford1995}. The presence of Alfv\'en waves \citep{Federman1996}, and the occurrence of turbulent dissipation \citep[e.g]{Godard2009} still remain viable mechanisms in diffuse clouds.
An attempt to provide a model which is able to reproduce the observed abundances has been proposed by \citet{Falgarone2010} exploring turbulent dissipation regions (TDR) in which dissipation of turbulent energy locally triggers a specific warm chemistry. This was found to well reproduce the results for the inner Galaxy conditions. 

The analysis of the present destruction reaction has also been carried out by several authors, who employed a variety of potential energy surfaces (PES) and tried to provide realistic estimates of the reaction rates for the destruction path of Eqn.(\ref{eq:reaction}) using an adiabatic fit of the various surfaces involved. Some have used the Negative Imaginary Potential (NIP) code within an infinite order sudden (IOS) approximation for the reaction dynamics \citep{Stoecklin2005}. More recent calculations which employ a newly computed reactive potential energy surface (RPES) were carried out by \citet{Warmbier} who used quasi-classical trajectories (QCT) and close-coupling (CC) quantum methods. 

We therefore think that it would be useful to employ the adiabatic PES approach once more to the depletion reaction (3), but using an accurate quantum method for angular momentum coupling, linking our findings to the recent study of \citet{Grozdanov2013} and also embed our final rates, after comparisons with the experiments, into a broader chemical network modelling the CH$^+$ evolution in dark clouds. To these ends, we decided to employ the recent RPES of \citet{Warmbier} and to use our recently developed NIP approach to quantum reactive  dynamics \citep{Tacconi2011} to study the depletion reaction of Eqn.(\ref{eq:reaction}), where the quantum dynamics was treated within the coupled-states (CS) approach. The use of the NIP method, originally suggested by \citet{Baer1990}, had been proved already to be quite realistic when dealing with the many channels which are usually dynamically coupled in ionic reactions \citep{Bovino2011b}. We shall specifically show below that the present results turn out to provide the best overall agreement with the existing experiments over a broad range of temperatures. 

The following Section briefly describes the main features of the employed RPES, together with an outline of our computational method for yielding reaction cross sections and rates. Section \ref{sect:rate_coefficients} presents our computed quantities, their low-T behaviour vis-\'a-vis changes in the RPES features, and further carries out an extensive comparison with a very recent computational modelling of non-adiabatic effects \citep{Grozdanov2013} together with a detailed analysis in relation to existing experiments. Section \ref{sect:modelling_ISM} describes the effects of our computed rates on realistic evolutionary models which include the CH$^+$ network of reactions, while the present conclusions are given in Section \ref{sect:conclusions}.

%*******************************************
\section{The quantum dynamics}
\begin{figure}
\begin{center}
\includegraphics[width=.45\textwidth]{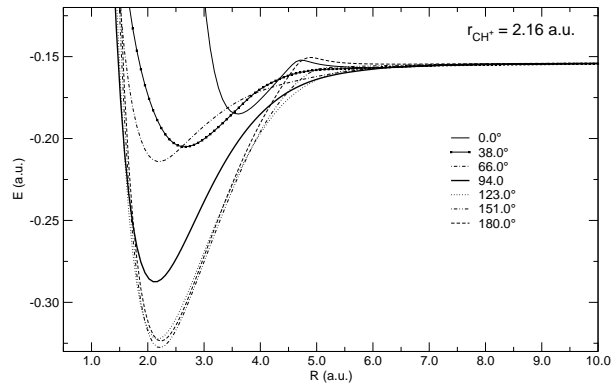}
\caption{Cuts of the PES at different angles for the CH$^+$ molecule at its equilibrium distance. }\label{figure1}	
\end{center}
\end{figure}

The CH$_2^+$ reactive system is characterised by the presence of a conical intersection between the $^2\Sigma^+$ and the $^2\Pi$ PES \citep{Stoecklin2005,Halvick2006,Warmbier}. The latter becomes avoided crossing for near-linear, bend configurations and suddenly disappears when the molecule is clearly bent. This behaviour has been discussed in great detail in the above studies and will not be repeated here. By performing an adiabatization of the relevant RPES one can obtain a smoother description of the lower, approximately adiabatic RPES which we have employed in the present work, as also done in all the existing previous studies. The resulting angular shape of this single RPES could be seen from the data in Fig.\ref{figure1}, where different cuts are shown for the CH$^+$ molecule at its equilibrium geometry. We note there that at angles of 0$^{\circ}$ or 180$^{\circ}$ (linear configurations) a small barrier occurs around 5 a.u., obviously depicting the presence of the avoided crossing. Therefore, the fit of the RPES should be carefully checked in those regions. In our calculations we employ the RPES calculated by \citet{Warmbier} constructed following a modified ansatz of \citet{Braams,Braams2008} and \citet{murrell1984}. This RPES is better in terms of accuracy than the previous work of \citet{Stoecklin2005} for two main reasons: (i) the number of points used to construct it is 16259 to be compared with 3291 used by the earlier authors, and (ii) the global root-mean-square found here is 15.5 meV to be compared with the 59.4 meV value of \citet{Stoecklin2005}. It should be added that, in both cases similar results for the two fragments (CH$^+$ and H$_2$) diatomic curve are provided,  also in good agreement with experiments. Furthermore, the RPES of \citet{Warmbier} provides a fitting which works better in the complex's region (CH$_2^+$), the latter being very important in terms of reactive scattering studies involving ionic species since they indeed proceed via the complex formation regions. An additional set of polynomials has been included for each intersection: as stated by the authors this approach is very sensitive to the chosen parametrisation but offers a more accurate description of the conical intersections when compared to the global accuracy of the whole potential fit \citep{Warmbier}. 

To describe the dynamics of the reaction in Eqn.(\ref{eq:reaction}), due to the features outlined above, is computationally very demanding since it should correctly be solved by taking into account the presence of at least two different PESs.  In practice, all published calculations thus far have resorted to using a single adiabatic RPES, due to the generally low-temperature regimes which are indicated for the ISM environments, and therefore considered the lower portion only of the conical intersection region and of both reagents' and products' states.

A very recent publication \citep{Grozdanov2013} has analyzed within a modified statistical model the possible effects of the conical intersections at the collinear orientations and we shall discuss their findings below, showing them to be in line with our own model analysis of such effects, also detailed below.
%In most of the cases the latter effect is considered by introducing a barrier. This corrective terms is valid for collision energies below the lowest point of the crossing seam and neglecting the rate of transition between the two electronic states.

Furthermore, the earlier computational/theoretical data related to this reaction have been obtained by employing methods which use an asymptotic basis or some approximation in the dynamics. For instance the negative imaginary potential (NIP) calculations reported by \citet{Stoecklin2005} include the IOS approximation which is generally known to underestimate the reactive cross sections \citep{Huarte}. The subsequent work by \citet{Halvick2006} employed a quasi-classical trajectory (QCT) and phase-space theory (PST) methods which are expected to be inaccurate at very low temperatures, where the quantum mechanical effects are dominant. The results from \citet{Warmbier} try to use a more accurate RPES and more accurate dynamics. They have been obtained via the close-coupling (CC) ABC code \citep{Skouteris} and explored the first five rotational levels for the CH$^+$ molecule. 

Our approach is based on the NIP method introduced by \citet{Baer1990} and extended in our previous works 
(\citealt{Bovino2011a,Bovino2011b,Tacconi2011}, see these papers for the mathematical details) which made use of an additional potential term, $V_{NIP}$, aimed at absorbing the reactive flux. 
Because of the flux-absorbing effects coming from it, the resulting S-matrix is non-unitary and its default to unitarity yields the (state-to-all) reaction probability, as discussed in \citet{Tacconi2011}.
From the reaction probability one can in turn obtain the reactive cross section 
\begin{equation}\label{eq:sigma}
	\sigma(E) = \frac{\pi}{(2j + 1)k^2}\sum_J\sum_{\Omega}(2J + 1) P^{J\Omega}\,,
\end{equation}
evaluated from a given initial state ($\nu$, $j$) and summed over all the final roto-vibrational states of the products. In Eqn.(\ref{eq:sigma}) $E$ is the collision energy, $k^2$ is the wave vector, $J$ the total angular momentum, and $\Omega$ the projection of the rotational angular momentum along the Body Fixed (BF) axis.   
Once the reactive cross sections are obtained, the rate coefficients are computed by averaging the appropriate
cross sections over a Boltzmann distribution of relative velocities:
\begin{equation}
	\alpha(T) = \frac{1}{(k_BT)^2}\sqrt{\frac{8k_BT}{\pi\mu}}
	%\int_0^\infty\sigma(E)e^{-\frac{E}{k_BT}}E\,\dd E
\end{equation}
where $T$ is the gas temperature, $k_B$ is the Boltzmann constant and
$\mu$ the reduced mass of the system. 

It is important to note here that our NIP implementation is different from the one discussed by \citet{Stoecklin2005}, where an asymptotic basis for the roto-vibrational CH$^+$ states is used throughout the entire reactive domain. In addition, their calculations make the assumption of the IOS approximation \citep{Kouri} that treats the rotational motion as ``frozen'' during the collisions so that the final cross sections are evaluated at fixed Jacobi angles. In our approach an adiabatic basis set is employed instead, thus taking into account the physical effects by which interaction between the incoming atom and the molecule can adiabatically modify the reagents' rotovibrational states. The reactive scattering calculations are carried out in the body-fixed frame making use of the coupled-states coupling scheme \citep{Mcguire,Mcguire2} approximation which has been found to provide accurate results when compared with experiments: see the earlier works by \citet{Bovino2011a,Bovino2011b}.

%*******************************************
\section{Rate coefficients and low temperature behaviour}\label{sect:rate_coefficients}
\subsection{The measured and computed rates}
We carried out calculations for the reaction in Eqn.(\ref{eq:reaction}) starting from the initial roto-vibrational level $\nu=0,j=0$ and for energies ranging from 10$^{-5}$ to 1.0 eV. Our results are thus providing final rates between 10-1000 K, which therefore extend the range of temperatures sampled by the earlier, accurate quantum calculations of \citet{Warmbier}.
All the parameters employed ensured a relative error of the calculated  reactive cross sections within 1\%. A basis set expansion of  800 functions has been used for the CH$^+$ reagent, leading
to an equal number of coupled equations. The molecular basis functions were expanded over a direct product of a Colbert-Miller discrete variable
representation (DVR) of 150 points (ranging from 0.35 $a_0$ to 15.0 $a_0$) and a set of 48 spherical harmonics.
The convergence over the total angular momentum values ($J$) has also been checked: we used a number of $J$ ranging  from 10 to 42 for the highest energy. Following the Baer criteria \citep{Baer1990} we have also obtained the following stable NIP parameters that have been employed in the calculations: $r_{\rm min} = 6.75\,a_0$, $r_{\rm max} =
10.25\,a_0$, and the NIP order $n = 2$ (see ref. \citealt{Tacconi2011} for further details).

\begin{figure}
\begin{center}
\includegraphics[width=.45\textwidth]{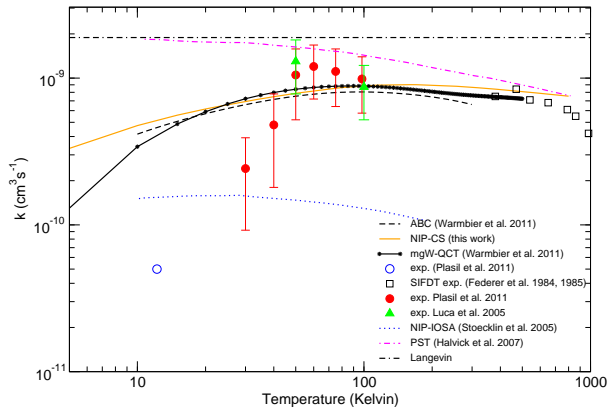}
\caption{Reaction rates for the destruction channel of Eqn.(\ref{eq:reaction}) as a function of the Temperature for different calculations and experiments. Colours online.}\label{figure3}
\end{center}	
\end{figure}

The final rates are shown in Fig.\ref{figure3} where they are compared with previous calculations and some experimental data. As can be seen from that figure, our results are both in good agreement with the accurate ABC-CC calculations which used the same RPES \citep{Warmbier} and with the experiment for temperature above 60 K \citep{Plasil2011}. It is worth noting here that for temperatures below 60 K the experiments by \citet{Plasil2011} are still seen to decrease with temperature faster than any of the available calculations. A possible explanation for such a behavior will be further discussed, while it is worth noting here that an independent set of additional measurements would be a welcomed addition to the available data. Other computed values obtained with the ABC and QCT methods \citep{Warmbier} employing the same RPES are, like us, in good agreement with the experiments, while the results from \citet{Stoecklin2005} and \citet{Halvick2006} appear to deviate from both set of data. This behaviour may be attributed to both their specific, model quantum approach (involving a series of approximations) as the poor accuracy of the PES involved. 

\begin{figure}
\begin{center}
\includegraphics[width=.45\textwidth]{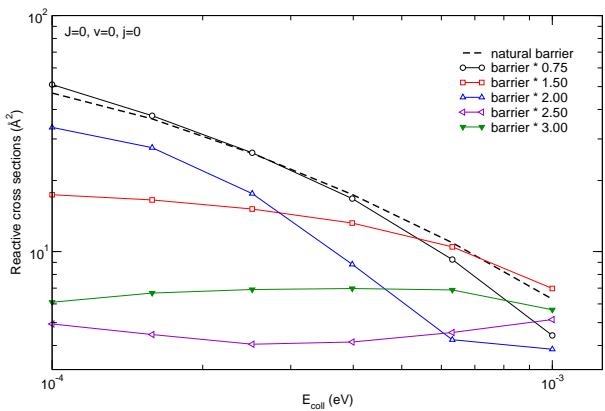}
\caption{Reactive cross sections as the function of the collision energy for different values of the $\alpha$ parameter. See text for the details. Colours online.}\label{figure2}	
\end{center}
\end{figure}

One of the surprising results reported by \citet{Plasil2011} is the sudden drop in the rates at temperature below 60 K seen by their experiments. Since the dynamics of the system has been explored with different computational methods, we thought that it would be interesting to further explore via a computational experiment another possible physical cause for such a feature, since the numerical diabatization artificially introduces a barrier as seen in our Fig.\ref{figure1}. More specifically, the presence of the barrier is due to conical intersections occurring for the collinear alignments of the reactants: H-H-C and H-C-H. A very interesting theoretical paper recently published on this subject \citep{Grozdanov2013} analyses the reactive behavior of such collinear configurations within a modified statistical treatment and finds that, by artificially suppressing the reactivity of the rotational states of CH$^+$ which dominate the reaction at the alignment channels, the final rates drop with temperature more rapidly, although not quite as the experiments. Their conclusions indicate that, if the presence of a barrier after diabatization were to be used to represent reactive flux losses into the other coupled surfaces, then a possible explanation of the existing experiments could be linked to the physical presence of such intersections. To explore this option we have carried out further model calculations in which we arbitrarily modify that barrier by using a disposable  artificial factor ($\alpha$) that can increase the barrier height along the crucially important linear configurations. This numerical experiment would reduce tunneling of the reagents into the inner region of the reactive lower surface, thereby mimicking the flow of dynamical flux into the upper surfaces excluded by diabatization. We have therefore carried out different calculations with different factor values to see how much changes in the barrier height at the collinear crossings can affect the final cross sections at low energies. In Fig.\ref{figure2} we report the results of our model for  different values of $\alpha$ ranging from 0.7 to 3.0. We show in this figure that an increase of the barrier leads to lower values of the cross sections at low temperatures, while for energies larger than 1 meV the differences are not so marked. It should be noted that in this numerical study we are testing only the contribution from the most important partial wave ($J=0$) and that when going to larger energies the higher $J$ contributions should be added. However, since the behaviour at very low energies is largely controlled by the s-wave, this test is a reasonable one and it allows to see the clear presence of a correlation between barrier and reactive probabilities.
This fairly simple numerical experiment is therefore telling us that diabatization is particularly important at the lower temperatures since the low-T alignment of the reactants would lead to additional flux losses into the other RPES's. Since the low temperature regime is exactly the region were the experiments are showing a marked drop in the size of the cross sections and rates, it follows that our numerical tests in that region have the dynamics efficiently modelled by observing instead the flux losses into different channels induced during the collinear encounters of reactants. The results of Fig.\ref{figure2} indicate that such effects are shown by our present study to be much less important as the temperature increases.
These findings are also relevant for the analysis, discussed in the next Section, of the CH$^+$ evolution within a large network of linked chemical processes in order to establish their bearing on the evolutionary abundances of the CH$^+$ molecule. We shall, in fact, show below that other chemical processes take over at the lower temperatures, thereby making less relevant the role of the present reaction. The latter, however, becomes very important again at the higher temperatures where the agreement between our calculations and the experiments, as shown in figure \ref{figure2}, is indeed quite good.

%In the end considering that two different reactive PES have been produced over the years and that different quantum reactive scattering have been performed leading to similar results, the deviation experiment provided deviation at low-T still remains a question point that needs additional confirmations.

%*******************************************
\section{The CH$^+$ chemistry in different ISM environments}\label{sect:modelling_ISM}

We have therefore analyzed the effects of the present rates on an ISM model which selects a specifically simplified chemistry based on H, He, C, and O elements, since the above components are the most abundant under the evolutionary conditions that we shall discuss below. Due to the ubiquitous presence of the CH$^+$ molecules we decided to model the evolution of the gas by exploring different environments: from a PDR to a self-shielded molecular cloud (MC). In order to do that,  we essentially vary the extinction coefficient $A_v$ and make it range from 0.2 to 10; we further select the temperature according to what is suggested by Sect.9.6 of \citet{Tielens2005}.
As indicated above, we employ a sub-network of the chemistry given in the UMIST database\footnote{http://www.udfa.net} \citep{UMIST2013} which includes 356 reactions and 34 species, the latter chemical species being listed in Tab.\ref{tab:species}. We then follow a model of a chemical network made up by simple radicals, and for this reason we do not include any of the more complex molecules such as the larger carbon chains. It is worth noting that some of the rate coefficients in the UMIST database have a limited temperature range of applicability which is mainly focussed into the cold regions ($\lesssim$300 K). We therefore extended such rate coefficients in order to cover a wider temperature range, following (whenever possible) the indications from the works where the rates had been evaluated.

We performed a series of standard one-zone models with different temperatures and different visual extinction coefficient values. In each model we take the temperature and density to be constant quantities during the evolution, while the chemical species are computed within a non-equilibrium scheme by using the chemical package \textsc{KROME}\footnote{Available for download from http://www.kromepackage.org} \citep{GrassiKROME} that employs a \textsc{DLSODES} solver \citep{Hindmarsh83, Hindmarsh2005}.
The initial conditions for the selected chemical species are given in Tab.\ref{tab:specie_init} as fractional abundances with respect to the total hydrogen number density $n_\mathrm{Htot}=2\times10^3$ cm$^{-3}$, while for the free electrons we adopt $n_\mathrm{e^-}=n_\mathrm{C^+}$ as initial value \citep{Cardelli1993,Meyer1998,Wakelam2008}. Finally, the cosmic rays ionisation rate is set to $1.3\times10^{-17}$ s$^{-1}$ in all the models, and we do not include any cooling or heating term during the 10$^8$ years evolution that we have sampled, in order to specifically assess the expected role of the chemical reactions included in our present network.

\begin{figure}
	\begin{center}
		\includegraphics[width=.45\textwidth]{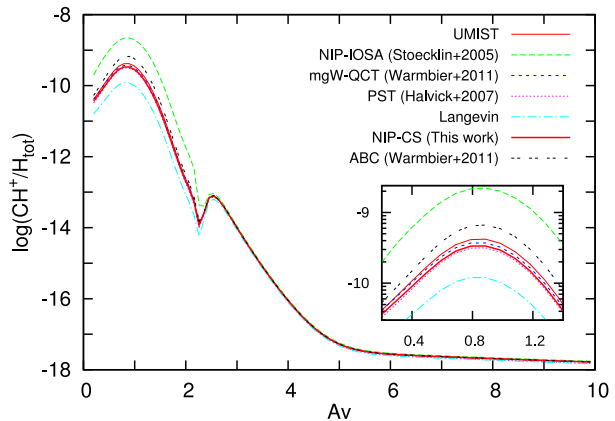}
		\caption{Fractional abundances of the CH$^+$ molecule normalised to the total density of the gas as a function of the visual extinction $A_v$. We present different profiles for different choices of the desctruction rates discussed in the present work (see text for details). The inset is an enlargement of the region between $A_v=0.3$ and $A_v=1.3$. Colours online.}\label{fig:Av_CHp}
	\end{center}
\end{figure}

The results of our calculations are shown in Fig.\ref{fig:Av_CHp} as the fractional abundance of the CH$^+$ molecule that varies with the visual extinction $A_v$, the later going from 0.2 to 10. This range of values influences the photoionisation and photodissociation rate coefficients that in the UMIST database have the form $k=\alpha\exp(-\gamma\,A_v)$ in units of s$^{-1}$, where $\alpha$ and $\gamma$ are two parameters that depend on the reaction considered. 

The amount of CH$^+$ shown in Fig.\ref{fig:Av_CHp} is markedly higher in the region with lower $A_v$, where the dominant formation path is controlled by the endothermic (4177 K) reaction \citep{Zanchet2013}
\begin{equation}\label{eqn:main_formation}
	\rm H_2 + \rm C^+ \rightarrow \rm{CH}^+ + \rm H\,,
\end{equation}
that is considered to be more efficient at higher temperatures (lower $A_v$). On the other hand, within the MC-like regions (higher $A_v$) where the temperature becomes lower the rate coefficient for Eqn.(\ref{eqn:main_formation}) is no longer efficient. Therefore, in the region around $A_v\approx2.2$ the profile presents a dip: from an anlysis of our calculations we found that this is due to the sudden increase of the H$_2$O abundance which depletes the C$^+$ abundance and hence reduces the efficiency of the formation by Eqn.(\ref{eqn:main_formation}). In the simplified network employed here this competing process is, in fact, triggered by the following endothermic reaction with oxygen
\begin{equation}\label{eqn:react_O}
	\rm H_2 + \rm O \rightarrow \rm{OH} + \rm H\,,
\end{equation}
that produces the OH molecule which in turn feeds the reaction
\begin{equation}\label{eqn:react_OH}
	\rm H_2 + \rm OH \rightarrow \rm{H_2O} + \rm H\,,
\end{equation}
and finally the H$_2$O produced is additionally further involved in the reactions
\begin{align}\label{eqn:react_H2O}
	\rm H_2O + \rm C^+ &\rightarrow \rm{HCO^+} + \rm H\nonumber\\
			   &\rightarrow \rm{HOC^+} + \rm H\,,
\end{align}
which consume the C$^+$ ion, the latter being the main source of CH$^+$ formation as indicated by Eqn.(\ref{eqn:main_formation}). To numerically test the above chain of reactions we found, in fact, that when we reduce the initial abundance of the atomic oxygen by a factor of two in our network, the dip is removed and the final amount of CH$^+$ increases by more than four orders of magnitudes, especially in the optically thick region ($A_v\gtrsim3$) corresponding to lower temperatures. This numerical experiments therefore confirms the suggested role of the above reactions within the chemistry of CH$^+$ and of CH$_2^+$.

We provide here two functional forms for our rate coefficent: one considered to be less accurate that follows the classical Kooij form
\begin{equation}\label{eq:ko}
	k(T) = \alpha\left(\frac{T}{300 \mathrm{K}}\right)^\beta\exp(-\gamma/T)\qquad \mathrm{cm^{3}\,s^{-1}}\,,
\end{equation}
with $\alpha = 8.72$ 10$^{-10}$, $\beta = -0.075$, $\gamma = 10.1866$, and $T$ in K, and another, more accurate fit (employed in these calculations) given by:
\begin{equation}\label{eq:koj}
	k(T) = a\sqrt{T} + bT + cT^{3/2} + dT^2\qquad \mathrm{cm^{3}\,s^{-1}}\,,
\end{equation}
where $a = 1.9336$ 10$^{-10}$, $b = -1.4423$ 10$^{-11}$, $c = 4.3965$ 10$^{13}$, $d = -4.8821$ 10$^{-15}$, and $T$ in K. Both fits are valid in the range of temperatures from 10 to 1000 K.

We report in Fig.\ref{fig:Av_CHp} the results from our evolutionary models which are obtained when using the various rate coefficients discussed in Sect.\ref{sect:rate_coefficients} and shown in Fig.\ref{figure3}, but also by additionally adding the Langevin value $k_{LV}=1.89\times10^{-9}$ cm$^3$ s$^{-1}$ which we obtained by following \citet{Levine1987}. Marked differences are found at lower $A_v$ when using either the NIP-IOS calculations or the Langevin rate coefficients. As expected, the latter rate is an upper limit for the reaction coefficient and therefore produces a smaller amount of CH$^+$, while the rate coefficient from \citet{Stoecklin2005}, which employs a dynamical approximation for the angular momentum coupling which has been found to underestimate the values of the cross sections, is also seen to generate larger abundances of CH$^+$. This behaviour is even more visible in the inset of Fig.\ref{fig:Av_CHp}. The evolutionary calculations which include our computed rates (labelled \emph{this work} in the figure) are found to be in good agreement with the UMIST, which is based on the experimental values of \citet{Plasil2011} above 60 K (see Sect.\ref{sect:rate_coefficients}). We are also, as expected, very close to the accurate quantum calculations of \citet{Warmbier}, labelled \emph{ABC} in the figure.

\begin{figure}
	\begin{center}
		\includegraphics[width=.48\textwidth]{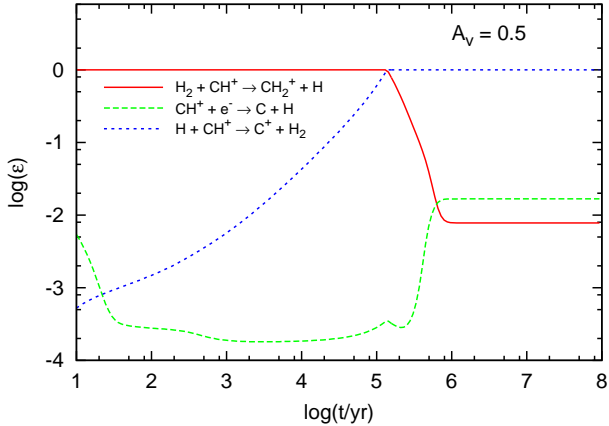}
		\caption{Evolution of the velocities of the main destruction reactions normalized to the most important one for $A_v=0.5$, see colured figure online.}\label{fig:destAv05}
	\end{center}
\end{figure}

\begin{figure}
	\begin{center}
		\includegraphics[width=.48\textwidth]{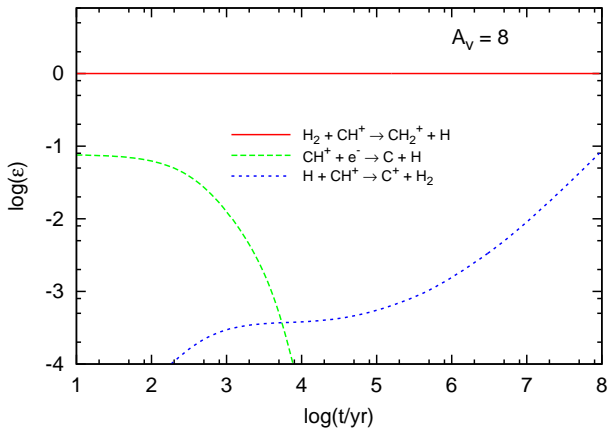}
		\caption{Same as Fig.\ref{fig:destAv05}, but for $A_v=8$. Colours online.}\label{fig:destAv80}
	\end{center}
\end{figure}

It is here useful to note that the differences at the lower $A_v$ do not originate here from the discrepancies of the rates at low temperature discussed in Sect.\ref{sect:rate_coefficients}, but rather depend on the high temperature conditions at these extinction coefficient values (see Fig.\ref{figure3}) where even the PST and the QCT produce reasonable results, although they are both failing at lower temperatures. To better understand this behaviour, we show the evolution of the velocities of the main reactions for the destruction of the CH$^+$ molecule at two different $A_v$ conditions: one representing a PDR-like environment ($A_v=0.5$, $T_\mathrm{gas}\approx800$ K), and another describing a MC-like ($A_v=8$, $T_\mathrm{gas}\approx30$ K) environment. The velocity $\epsilon_i$ of the \ith reaction is defined as 
\begin{equation}
	\epsilon_i=\frac{k_i\,n_{r1i}\,n_{r2i}}{\max\left(k\,n_{r1}\,n_{r2}\right)}\,,
\end{equation}
where $k_i$ is the rate coefficient, $n_{r1i}$ and $n_{r2i}$ are the abundances of the reactants and the denominator represents the velocity of the fastest destruction rate at a given time. 
These quantities are reported in Figs.\ref{fig:destAv05} and \ref{fig:destAv80}, for $A_v=0.5$ and $A_v=8$, respectively. In the optically thin regime, where the photodissociation of H$_2$ is more efficient, we note that the reaction 
\begin{equation}\label{eqn:R1}
	\rm H_2 + \rm{CH}^+ \rightarrow \rm{CH_2^+} + \rm H\,,
\end{equation}
dominates the CH$^+$ destruction, while, when the hydrogen becomes atomic, the channel of Eqn.(\ref{eq:reaction}) starts to be important.
The optically thick environment is dominated by the presence of H$_2$ during the whole evolution, and therefore the reaction of Eqn.(\ref{eqn:R1}) always plays a key role.
These plots clearly show that in the lower temperature region, where the computed rates are markedly different from one another (see Fig.\ref{figure3}), the present evolutionary results are seen to be only marginally influenced by such differences since this destruction channel becomes in fact negligible (see Fig.\ref{fig:destAv80}). Conversely, at the expected higher temperatures of the PDR-like region, the smaller differences between the rates arise because this new environment is now becoming richer in atomic hydrogen at the later temporal stages (see Fig.\ref{fig:destAv05}).
In addition, from Figs.\ref{fig:destAv05} and \ref{fig:destAv80} we note that, given our selected initial conditions and the sub-network employed, the dissociative recombination is not likely to be important for the destruction of the CH$^+$ molecule when compared to the other existing destruction channels. 

%%********* CH+ with Av ********

%*************** table all species ***************************
\begin{table}
	\caption{List of the species included in our ISM model. See text for details.}
	\begin{center}
		\begin{tabular}{llll}
			\hline
			Species &&&\\
			\hline
	e$^-$ &  H &  H$^-$ & H$^+$  \\
	He &  He$^+$ & C &  C$^-$  \\
	C$^+$ & O &  O$^-$ &  O$^+$  \\
	H$_2$ &  H$_2^+$ &  C$_2$ &  CO  \\
	CO$^+$ &  O$_2$ &  O$_2^+$ &  OH  \\
	OH$^+$ &  CH &  CH$^+$ &  HCO \\
	HCO$^+$ &  CH$_2$ &  CH$_2^+$ &  CH$_3$ \\ 
	CH$_3^+$ &  H$_2$O &  H$_2$O$^+$ & HOC$^+$ \\
	H$_3^+$ &  H$_3$O$^+$ & \\
			\hline
		\end{tabular}
	\end{center}
	\label{tab:species}
\end{table}

%*************** table initial ***************************
\begin{table}
	\caption{Initial fractions respect to $n_\mathrm{Htot}$.}
	\begin{center}
		\begin{tabular}{ll|ll}
			\hline
			Species & Fraction & Species & Fraction\\ 
			\hline
			H$_2$ & $0.50$ 	& He & $9.00\times10^{-2}$\\
			O & $2.56\times10^{-4}$ 	&	C$^+$ & $1.20\times10^{-4}$\\
			e$^-$ & see text\\
			\hline
		\end{tabular}
	\end{center}
	\label{tab:specie_init}
\end{table}

%%********* destruction Av 0.5 ********

%%********* destruction Av 8.0 ********

%---------------------------------------------------------------------------
\section{Conclusions}\label{sect:conclusions}
In the present work we have attempted to address various, separate aspects involving the chemistry of the CH$^+$ in the ISM, combining them together in order to gain a better understanding of its role as a chemical component in many diverse environments:
\begin{enumerate}
\item[(i)] we have carried out a new theoretical evaluation of the chemical path to its destruction via the chemical reaction with H atom discussed in the introduction as Eqn.(\ref{eq:reaction});
\item[(ii)] we have employed the reaction rates given by our calculations within different models of its evolution associated with either a PDR-like, high-temperature environment or with an MC-like model at lower temperatures.
\item[(iii)] we have carried out numerical modelling tests on the features of the employed RPES in the region of the conical intersections in order to show that reactive flux losses into the other coupled RPES's could be relevant at the very-low temperatures, and therefore could be used to partly explain the experimentally observed drops in rate values as T decreases.
\item[(iv)] the present evolutionary study of a large set of connected chemical reactions clearly shows that the above, low-T regime of the reaction (3) is marginally affecting the CH$^+$ presence in the astrophysical environments that we have modelled, since under those conditions other chemical reactions begin to play the role of destroying that species.
\end{enumerate}
In particular, the new analysis of CH$^+$ depletion channel (Eqn. \ref{eq:reaction}) stems from the quantum calculations of the reactive cross section using an accurate RPES already available from the recent literature \citep{Warmbier} and employing a quantum reactive method, the NIP approach used by us before \citep{Bovino2011a}, with a CS treatment of the dynamical coupling during the evolution from reagents to products. A detailed comparison of our new findings with the available experiments and with all recent calculations of the same reaction (see details in Sect.\ref{sect:rate_coefficients} and Fig.\ref{figure3}) turns out to indicate that our calculated rates are in very good agreement with experiments and also with the best available quantum calculations for the same depletion reaction over the range of astrophysically relevant temperatures. Furthermore, given the differences shown by all existing calculations (old and new) with the experiments for the low-T data \citep{Plasil2011}, and in order to check a possible physical cause  of the rate values reduction at the low temperatures (see Fig.\ref{figure3}), we have also carried out a numerical experiment whereby the barrier in the product region, caused by the avoided crossings at the different orientations of the regents (see Fig.\ref{figure1}), is artificially modified in height and the ensuing reactive probabilities analysed. We thus found that any role of such a barrier in driving the destruction reaction is very relevant in the low-T region, while becoming negligible as the temperature is increased. We could therefore suggest from such computations that the low-T behaviour of the reaction rates for this system are indeed sensitive to specific details of the adiabatisation of a multiple-RPES reaction. It will translate, in the experiments, into the possible role of non-adiabatic effects which would be very delicate to measure reliably in that region. Additionally, our detailed modelling of two distinct astrophysical environments, within which the reaction (3) is linked to a broader range of chemical processes, has shown that the low-T regimes of the latter reaction play a marginal role for describing its evolution, which it is instead dominated by the high-T behaviour where our quantum calculations indeed match existing experiments.

The second aspect of the CH$^+$ chemistry that we have analysed in some detail, as described in the previous Sect.\ref{sect:modelling_ISM}, has been the monitoring of its evolution with different visual extinction values ($A_v$) in order to model either the high temperature conditions of a PDR environment with intense photon flux (low $A_v$ values) or the low temperature conditions of a photon-thick environment for a MC. In both cases, at each $A_v$ value the temperature has been kept constant and the additional presence of other radical atoms and ions (together with all their interconnected reactions) has been included in our simulations by accessing the database UMIST to assemble a sub-network excluding the presence of less important paths. The evolution has been followed through the publicly available package \textsc{KROME} \citep{GrassiKROME} for $10^8$ years and the results were collected in Figs.\ref{fig:Av_CHp} through \ref{fig:destAv80}.   

From those data we were able to understand that the role of the present destruction reaction is an important one in the PDR conditions while, as the temperature and the photon flux decrease, other destruction reactions take over and therefore even large changes in the destruction rates for the present reaction play an overall minor role.

In addition, we should consider that our calculated rate coefficient could play a key role at lower temperatures in the H-dominated enviroment like the cold neutral medium (CNM) regions, where the differences in the CH$^+$ evolution is supposed to be led by the destruction reaction studied in this work.

In conclusion, we have shown that using accurate quantum methods for the destruction reaction of CH$^+$ with H does yield very reliable reaction rates, which are in accord with existing experiments and which further appear to be important at high-T values, while becoming less crucial in modelling the CH$^+$ abundances in the photon-poor, low-T ISM regions. The present computational rates are also found to agree well with those included already within the UMIST database since the latter rates are indeed based on the experimental data above 60 K \citep{Plasil2011}.

\section*{Acknowledgements}
S.B. thanks for funding through the DFG priority programme `The Physics of the Interstellar Medium' (project SCHL 1964/1-1).
One of us (T.G.) also thanks CINECA consortium of the awarding of financial support while the present research was carried out.
We are very grateful to Robert Warmbier for having provided the PES and shared his results and to D.R.G Schleicher for having read this Manuscript and for his useful suggestions on the presentation of our results.

\bibliographystyle{mn2e}

%\section*{Acknowledgements}
%S.B. and D.R.G.S. thank for funding through the DFG priority program `The Physics of the Interstellar Medium' (projects SCHL 1964/1-1). D.R.G.S. thanks for funding via the SFB 963/1 on "Astrophysical Flow Instabilities and Turbulence". 

%
%\bibliographystyle{mn2e}      
%\bibliography{mybib_new} 

%---------------------------------------------------------------------------
%---------------------------------------------------------------------------
%---------------------------------------------------------------------------

\bsp

\label{lastpage}

\end{document}